\documentclass[11pt,titlepage]{article}

\usepackage{epsfig}
\usepackage{amsmath}
\usepackage{amssymb}
\usepackage{array}
\usepackage{setspace}
\usepackage[round,numbers,sort&compress]{natbib}

\title{Ensemble based convergence analysis of biomolecular trajectories}
\author{Edward Lyman
        and Daniel M. Zuckerman\thanks{
           Corresponding author.  Address:
           University of Pittsburgh,
           3079 BST3, 3501 Fifth Ave.
           Pittsburgh, PA~15213, U.S.A.,
           Tel.:~(412)648-3335, email: dmz@ccbb.pitt.edu}\\ 
        Dept. of Computational biology, School of Medicine\\ 
	and Dept. of Environmental and Occupational Health, Graduate School of Public Health,\\ 
        University of Pittsburgh, Pittsburgh, PA~15213}

\date{\today}

\begin{document}
\maketitle
\begin{abstract}
Assessing the convergence of a biomolecular simulation is an essential 
part of any careful computational investigation, because many 
fundamental aspects of molecular behavior 
depend on the relative populations 
of different conformers.
Here we present a physically intuitive method to self-consistently 
assess the convergence of trajectories generated by molecular dynamics 
and related methods. 
Our approach reports directly and systematically on the 
structural diversity of a simulation trajectory. Straightforward 
clustering and classification steps are the key ingredients, 
allowing the approach to be trivially applied to systems of any size. 
Our initial study on met-enkephalin strongly suggests that even fairly 
long trajectories ($\sim 50$ nsec) may not be converged for this 
small---but highly flexible---system.
\end{abstract}
\emph{Key words:} Convergence; met-enkephalin; efficiency; structural histogram 
\clearpage
\section{Introduction}
Conformational fluctuations are essential to the functions of proteins, whether 
they are motor proteins\cite{svoboda-motor-pnas94}, enzymes\cite{gordon-glu-biochem2000,kern-cypa-science02}, 
signalling proteins\cite{ikura-camapo-nature95,kern-allosteric-science01,cam-rdc-pnas2004}, or almost any other kind. 
Different experiments have enabled observation of protein fluctuations over a huge range of timescales, 
from picoseconds\cite{anfinrud-science03} to microseconds\cite{kern-allosteric-science01} 
to milliseconds\cite{kern-cypa-science02,cam-rdc-pnas2004,ubq-nmr-jmb05} to seconds and longer\cite{vendruscolo-hx-jacs2003}.   
 
Naturally, simulations aim to observe conformational fluctuations as 
well. A gap remains, however, between the timescale 
of many biologically important motions ($\mu$sec--sec), and that 
accessible to atomically detailed simulation (nsec). 
To put it another way, some problems are simply not possible to study 
computationally, since it is so far impossible to run a simulation which 
is ``long-enough.'' 

For those problems which are at the very edge of being 
feasible, we would like to know whether we have indeed sampled enough 
to draw quantitative conclusions. 
These problems include the calculation of free energies of 
binding\cite{pearlman-jmedchem05,shirts-binding-jcp05}, 
ab initio protein 
folding\cite{baker-fold-science05,simmerling-trpcage-jacs02}, and simulation of flexible 
peptides\cite{freed-metenk-solv} and conformational changes\cite{zuckerman-calmod}.

Convergence assessment is also crucial for rigorous tests 
of simulation protocols and empirical force fields---see, e.\ g.\cite{freed-uavaa}. Many algorithms propose to 
improve the sampling of conformation space, but quantitative 
estimation of this type of efficiency is difficult---except in simple cases\cite{coolwalk}.
In the case of force field validation, it is important to know whether 
systematic errors are a consequence of the force field, or are due to undersampling.

The observed convergence of a simulation depends on how convergence is defined and 
measured. It is therefore important to consider what sort of quantity is to 
be calculated from the simulation, and choose an appropriate way to assess 
the adequacy of the simulation trajectory (or trajectories). Many 
relatively simple methods are commonly used, such as 
measuring distance from the starting structure as a function of simulation time, 
and calculation of various autocorrelation functions\cite{freed-uavaa,duan-efficiency-jcp05}. 
Other, more sophisticated 
methods are based on principal components\cite{hess-pre02,garcia-metenk-prot02} 
or calculation of energy-based ergodic measures\cite{thirumalai-jacs93}. 

Many applications, however, require a thorough and equilibrated sampling of the 
space of \emph{structures}. All of the methods just listed are only related 
indirectly to structural sampling. There are many examples of groups 
of structures which are very close in energy, but very dissimiliar structurally. 
In such cases, we might expect energy--based methods to be insensitive to the 
relative populations of the different structural groups. It is therefore of 
interest to develop methods which are 
more directly related to the sampling of different structures, and see how 
such methods compare to more traditional techniques. 

Daura et.\ al.\ previously considered convergence assessment by 
\emph{counting} structural clusters, based upon a cutoff 
in the RMSD metric\cite{vangun-cluster99,vangun-cluster02}. 
The authors assess the 
convergence of a simulation by considering the number of 
clusters as a function of time. Convergence is deemed sufficient when 
the curve plateaus. This is surely a better measure than simpler, 
historically used methods, such as RMSD from the starting structure or 
the running average energy. However, it is worth noting that long after 
the curve of number of clusters vs.\ time plateaus, the \emph{relative populations} 
of the clusters may still be changing. 
Indeed, an important conformational 
substate which has been visited just once will appear as a cluster, 
but its relative population will certainly not have equilibrated.

The method of Daura et.\ al. also suffers from the need to store the entire 
matrix of pairwise distances. For a trajectory of length $N$, the memory 
needed scales as $N^{2}$, rendering the method impractical for long 
trajectories. At least two groups have developed methods which 
rely on nonhierarchical clustering schemes, and therefore require 
memory which is only linear in $N$. Karpen et.\ al. developed a 
method which optimizes 
the clusters based on distance from the cluster 
center\cite{brooks-art2clust-biochem93}, with distances 
measured in dihedral angle space. Elmer and Pande 
have optimized clusters subject to a constraint on the number of 
clusters\cite{pande-drms}, with distance defined by the 
atom-atom distance root mean square 
deviation\cite{ooi-drms-72,levitt-coarsefold-jmb76}. 

In this paper, we address systematically the measurement of sampling quality. 
Our method classifies (or bins) a trajectory 
based upon the ``distances'' 
between a set of reference structures and each structure in the trajectory. 
Our method is unique in that it not only builds clusters of structures, 
it also compares the cluster 
populations. By comparing different fragments of the trajectory 
to one another, convergence of the simulation is judged 
by the relative populations of the clusters. We believe  
the key to assessing convergence is tracking relative bin populations. 
Our approach can be directly applied to comparing the efficiency of 
different sampling methods. 
%

In the next section, we present a detailed description of the algorithm 
and discuss possible choices of metric. We then demonstrate the method on 
simulations of met-enkephalin, a structurally diverse peptide. 
%
%
\section{Theory and methods}
We will evaluate sampling by comparing ``structural histograms'', 
described below. These histograms provide a fingerprint 
of the conformation space sampled by a protein, by projecting 
a trajectory onto a set of bins based on distinct reference structures. Comparing 
histograms for different pieces of a trajectory (or for 
two different trajectories), projected 
onto the same set of reference structures, provides a very 
sensitive measure of convergence. Not only are we comparing 
how broadly has each trajectory sampled conformation space, but 
also how frequently each substate has been visited.        
\subsection{Histogram construction\label{algorithm}}
We generate the set of reference structures and corresponding 
histogram from a trajectory in the following simple way (our choice 
for measuring conformational distance will be discussed below):

\parbox{5in}{ 
(i) A cutoff distance $d_{c}$ is defined.\\
(ii) A structure $S_{1}$ is picked at random from the trajectory.\\ 
(iii) $S_{1}$ and all structures less than $d_{c}$ from $S_{1}$ are 
removed from the trajectory.\\
(iv) Repeat (ii) and (iii) until every structure in the trajectory 
is clustered, generating a set 
$\{S_{i}\}$ of reference structures, with $i=1,2,...$.\\
(v) The set $\{S_{i}\}$ of reference structures is then used to build a histogram:
every structure in the trajectory is classified according to its \emph{nearest}
reference structure. Note that this \emph{classification} step generates a unique 
histogram for a given set of reference structures---unlike the simple clustering 
which is generated in step (iii).\\
}

Such a partitioning guarantees a set of clusters whose centers are at 
least $d_{c}$ apart. Furthermore, for a trajectory of $N$ frames, 
the number of reference structures, $M$, 
and therefore the memory needed to store the resulting $M\times N$ 
matrix of distances, is controlled by $d_{c}$. For 
physically reasonable cutoffs (e.g., $d_{c} \gtrsim 1$ \AA\ RMSD), the number of 
reference structures is at least an order of magnitude smaller than the 
number of frames in the trajectory. The memory requirements are therefore 
manageble, since the computation of pairwise distances scales as $N\log N$.  

There is nothing in principle which prevents the use of a more carefully chosen 
set of reference structures with our classification scheme. 
For example, we may consider a set of structures which correspond to 
minima of the potential energy surface. The cutoff might then be chosen 
to be the smallest observed distance between any pair of the minimum energy 
structures, and the set of reference structures so determined could be augmented 
by the random selection of more reference structures in order to span the whole trajectory. 

However, 
we expect that the purely random selection used here will naturally include 
the lowest free energy substates, since these are the most populated. In either case, 
\emph{any} set of reference structures defines a unique histogram for any 
trajectory.   

%
\subsection{Trajectory Analysis}
Once we have a set of reference structures, we may easily compare 
\emph{two different trajectories} 
classified by the same \emph{set of reference structures}, by 
comparing the populations of the various bins as observed in the 
two trajectories: 
given a (normalized) population $p_{i}(1)$ for 
cluster $i$ in the first trajectory, and $p_{i}(2)$ in the second, 
the difference in the populations $\Delta P_{i} = |p_{i}(1) - p_{i}(2)|$ 
measures the convergence of substate $i$'s population between the two trajectories. 
 
Note that the ``two'' trajectories just discussed may be two different 
pieces of the same simulation. In this way, we may self-consistently 
assess the convergence of a continuous simulation, by looking to see whether the 
relative populations of the most populated substates are changing 
with time. Of course, this cannot answer affirmatively that a simulation 
has converged (no method can do so); however, it may answer negatively. 
In fact, 
we will see later that our method indicates that structural convergence may be much 
slower than previously appreciated.

Our approach should also be applicable to some types of non-continuous trajectories, 
such as those generated by multiple starts (e.\ g.,\cite{caves-multirun98}) or 
parallel exchange protocols (e.\ g., \cite{hansmann-pt97,okamoto-repex-md}). For 
multiple independent trajectories, one can compare the two histograms generated 
from (i) the first halves and (ii) the second halves of all simulations. If 
converged, these two histograms should agree. One could also compare histograms 
generated by grouping entire trajectories into distinct sets. For a parallel exchange 
simulation, where the ensemble is built from a set of continuous trajectories, 
histograms from the first and second halves of the simulation can be compared.

The comparison of histograms clearly will \emph{not} be appropriate when 
ensembles are generated in a fully decorrelated way. For instance, starting from 
a single long trajectory, one could generate two ensembles by randomly selecting structures, 
or perhaps by selecting structures at two different fixed time intervals. So long as the number 
of structures in each ensemble greatly exceeds the number of reference structures used for 
classification, it is hard to see how such histograms could be significantly different. In such 
cases, dynamical correlations have been explicitly discarded, and the histograms can only differ 
statistically.   
\subsection{Structural Metrics}
Many different metrics have been used to measure distance between conformations. 
The choice depends on both physical and mathematical 
considerations. For example, dihedral angle based metrics are well-suited to 
capture local structural information\cite{brooks-art2clust-biochem93}, but are not 
sensitive to more global rearrangements of the molecule.  
Least-squares superposition 
followed by calculation of the average positional fluctuation per atom 
(RMSD) is quite popular, but the problem of optimizing the superposition 
can be both subtle and time-consuming for large, multi-domain 
proteins\cite{snyder-clusterprec-proteins05}. In addition, RMSD does not satisfy a triangle 
inequality\cite{crippen-antifunnel-prot98}. This is not an issue for the algorithm 
presented here, but is a consideration for more sophisticated clustering 
methods\cite{pande-drms}. We will use RMSD to measure distance here, though 
we note that 
``distance root mean square deviation'' (drms) 
(or sometimes, ``distance matrix error'')\cite{ooi-drms-72,levitt-coarsefold-jmb76} 
may be appropriate when RMSD is not.

Labelling two structures by $a$ and $b$, 
the traditional  root mean square deviation (RMSD) 
is defined to be the minimum of the root mean square average of 
interatomic distances over all possible translations and rotations 
of $\mathbf{x}^{b}$---namely, 
\begin{equation}
\text{RMSD}(a,b) = \underset{\mathbf{x}^{b}}{\text{min}}\left\{\sqrt{\frac{1}{N}\sum^{N}_{j=1}||\mathbf{x}_{j}^{a} - \mathbf{x}_{j}^{b}||^{2}}\right\},
\label{rmsd}
\end{equation}
where $N$ is the number of atoms and $\mathbf{x}_j$ is the position 
of atom $j$.

It is clear that the choice of $d_{c}$, together with the choice of metric, 
determines the resolution of the histogram. Reducing $d_{c}$ increases the 
number of reference structures, and reduces the size of the bins. How is $d_{c}$ 
chosen? There is no general answer, and a suitable cutoff will depend on the problem under 
investigation. 

The typical RMSD between a pair of structures will depend on 
the size of the molecule, its flexibility, and the conditions of the simulation. 
If the magnitude of some important conformational change is 
known in advance, then this information will guide the selection of an 
appropriate cutoff. If not, then a series of histograms should be 
constructed at several values of $d_{c}$. The behavior of the histogram as a function 
of $d_{c}$ will give a sense of the appropriate value, as we will see below.

\section{Results}
We have tested our classification algorithm on implicitly solvated 
met-enkephalin, a 
pentapeptide neurotransmitter. By focusing first on a small peptide, 
we aim to develop the methodology on a system which may 
be thoroughly sampled and analyzed by standard techniques. 

The trajectories analyzed in this section were generated by Langevin 
dynamics simulations, as implemented in the Tinker v.\ 4.2.2 simulation 
package\cite{tinker}. The temperature was $298$ K, the friction constant was 
$5$ ps$^{-1}$, and solvation was treated by the GB/SA method
\cite{still-gbsa}. Two $100$ nsec trajectories were generated, each 
starting from the PDB structure $1$plw, model $1$. The trajectories 
will be referred to as plw-a and plw-b. Coordinates were written every 
$10$ psec, for a total of $10^{4}$ frames per trajectory.


\subsection{Previous methods: RMSD analysis and cluster counting}
\begin{figure}[tp]
\epsfig{file=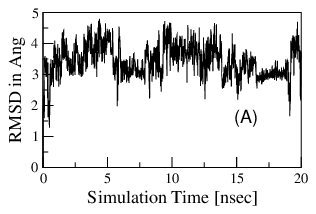 ,width = 0.50\columnwidth}
\epsfig{file=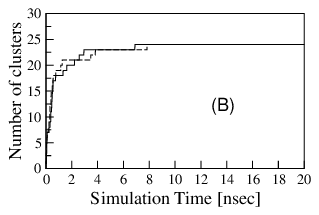 ,width = 0.50\columnwidth}
\caption{
}
\label{fig2}
\end{figure}

An often used indicator of equilibration is the RMSD from the 
starting structure (see Fig.\ \ref{fig2}A). Such plots are motivated by 
the recognition that the starting structure (e.g., a crystal structure) 
may not be representative of the protein under the simulation 
conditions---solvent, force field, and temperature. This is the case in 
Fig.\ \ref{fig2}A---the computation was performed with an implicit water model, while 
the experimental structure was determined in the presence of bicelles\cite{metenk-bicelle-nmr}. 
The system fails to settle 
down to a relatively constant distance from the starting structure---rather, 
it is moving between various substates, some nearer and some farther from 
the starting structure. While this is not surprising for a peptide renowned 
for its floppy character, it also indicates that this method cannot 
determine when the peptide simulation has converged. Indeed, Fig.\ \ref{fig2}A 
can tell us little about the convergence of the simulation, only that it 
spends most of its time more than $2.0$ \AA\ from the starting structure.  

A perhaps better indication of equilibration is provided by Fig.\ \ref{fig2}B, 
in which we have used the method of Daura, et.\ al\cite{vangun-cluster99}, 
albeit with clusters built by the procedure described in Sec.\ \ref{algorithm}. 
The premise is that convergence is achieved when the number of clusters 
no longer increases, as this means that the simulation has visited every 
substate.
This analysis suggests that convergence is observed by about $7$ nsec, and 
the curve has the comforting appearance of saturation. 
However, Fig.\ \ref{fig2}B is insensitive to the 
\emph{relative populations} of the clusters. To illustrate the problem, 
consider a simple potential, with two smooth wells separated by a high 
barrier. By simple cluster counting, a simulation will be converged as soon as it 
has crossed the barrier once. It is clear, however, that many crossings will 
be required before the populations of the two states have equilibrated. 
We will address this question  
using our ensemble-based method. We find, in fact, that the \emph{relative 
populations} of the clusters continue to change, long after their number 
has equilibrated.     
%
\subsection{Ensemble--based assessment of trajectories}
The use of our systematic approach is much more revealing. 
We first discuss the selection of an appropriate cutoff. We then demonstrate 
two different applications of the ensemble based comparison of 
trajectories---a comparison between a trajectory and a ``gold standard'' 
ensemble, and a self-consistent convergence analysis of a single trajectory.
\subsubsection{Reference structure generation and cutoff selection}
A compound trajectory was formed from trajectories plw-a and plw-b, 
by discarding 
the first $1$ nsec of each trajectory and concatenating the two into 
a single, $198$ nsec trajectory (plw-ab). We then generated a set of reference 
structures for the compound trajectory, as described 
earlier: a structure is picked at random, and it is temporarily 
discarded along with every structure within a predefined 
cutoff distance, $d_{c}$. The process is repeated on the remaining 
structures until the trajectory has been exhausted. The result is a set of 
reference structures which are separated from one another by at least 
the pre-defined cutoff distance. Lowering the cutoff 
(making the reference structures more similiar) increases the 
resolution of the clustering, and increases the number reference 
structures (see Table \ref{cutoff}). While RMSD 
is system-size dependent\cite{crippen-rmsd-jmb94}, 
for a particular system the cutoff 
defines a resolution. 

\begin{figure}
\epsfig{file=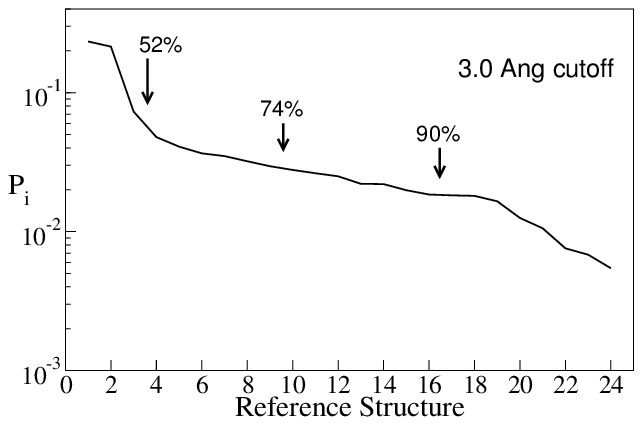,width = 0.50\columnwidth}
\epsfig{file=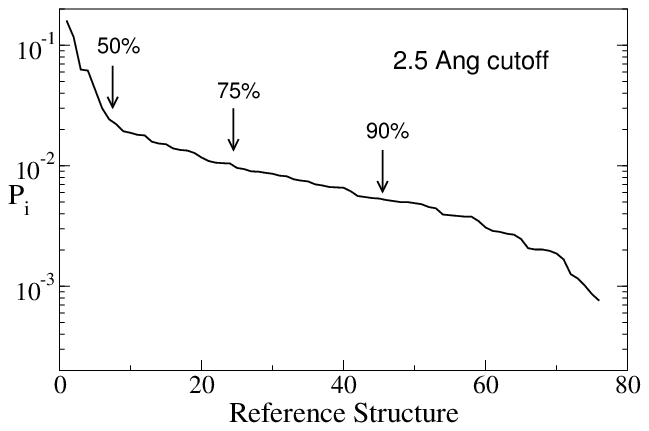,width = 0.50\columnwidth}
\epsfig{file=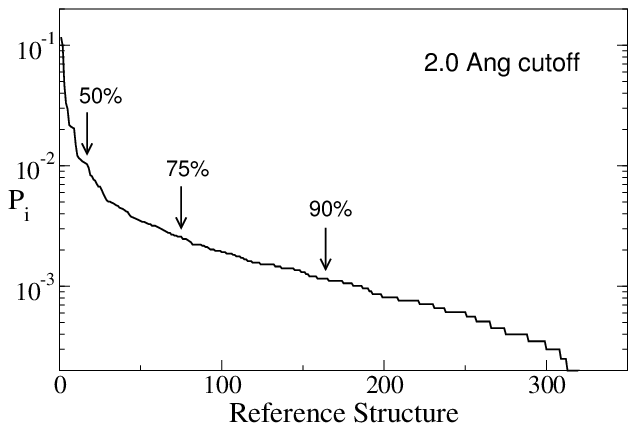,width = 0.50\columnwidth}
\epsfig{file=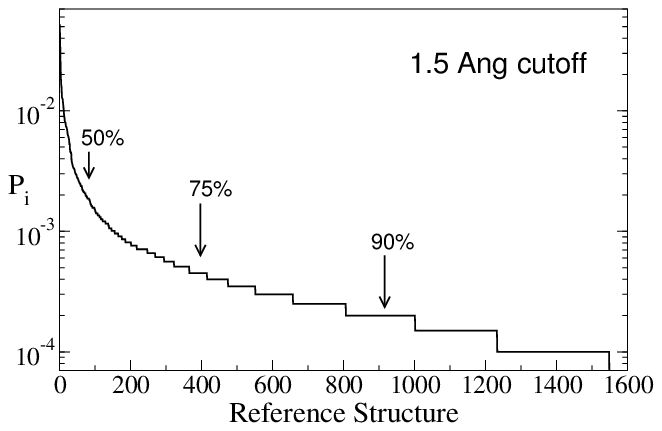,width = 0.50\columnwidth}
\caption{
}
\label{diffcutoffs}
\end{figure}

A histogram is then constructed by grouping each frame in the trajectory with 
its nearest reference structure.
The dependence of the histogram on $d_{c}$ is shown in Fig.\ \ref{diffcutoffs}. 
With $d_{c}=3.0$ \AA\, the first three bins already account for more than 
$50$\% of the total population. It might be expected that such a coarse description of 
the ensemble may not be particularly informative---however, we will see  
in the next sections that this level is already sufficient to make 
powerful statements about convergence. 

Lowering the cutoff, the general features of the histogram remain 
unchanged: a steep slope initially, which accounts for half of the total 
population, followed by a flatter region. In each case, most ($90$\%) of the population 
is accounted for by approximately half of all the reference structures. However, a 
closer inspection reveals that the fraction of bins required to account for the 
noted percentages of population ($50$, $75$, and $90$\%) is decreasing with 
the cutoff. For example, for $d_{c}=3.0$ \AA\, $16$ of $24$ bins account for 
$90$\% of the trajectory, while for $d_{c}=2.0$ \AA\, $164$ of $331$ bins account for
$90$\% of the trajectory. It should be mentioned, however, that this difference between 
the $d_{c}=2.0$ \AA\ and $d_{c}=1.5$ \AA\ histograms is so small as to be insignificant. 

Although it seems obvious that the most revealing cutoff will be system-specific, our 
histograms are more robust than they first appear. Because reference structures are chosen 
arbitrarily, the divisions between bins will not reflect basins of the landscape. 
In other words many, if not most, bins can be expected to include a number of full and partial 
local basins. Thus a lack of convergence in a ``macroscopic'' bin, at least in principle, can 
report on more local, microscopic states. Further, because our approach is so inexpensive 
compared to the simulation itself, more than one binning of configuration space can (should) 
be considered: see Sec.\ \ref{convergencesec} and Fig.\ \ref{fig4}.  

Based upon the series of histograms in Fig.\ \ref{diffcutoffs}, we continued our 
study of met-enkephalin based upon $d_{c}=3.0$ \AA. At this level of resolution, 
the main features of the histogram are already present, while the number of reference 
structures is small enough to make the computation quite inexpensive. We shall 
see that $d_{c}=3.0$ \AA\ provides sufficient resolution to 
investigate the convergence properties of our simulation.

Though we do not pursue it here, we note that the tail of the distribution---where 
half of all the bins account for only $10$\% of the population---might contain 
some very interesting structures. Indeed, at the very end of the tail are found 
bins which sometimes contain a single structure. Might some of these low population bins 
represent transition states? For now, we set this question aside, and focus instead 
on convergence assessment.
%
\subsubsection{Comparing trajectories to a ``gold standard'' ensemble}
In some applications, we want to compare a trajectory to a ``gold standard'' ensemble. 
For example, the gold standard might be the ensemble sampled by a long molecular dynamics 
simulation, and we may wish to check the ensemble produced by a new simulation protocol 
against the long molecular dynamics trajectory.

\begin{figure}[tp]
\epsfig{file=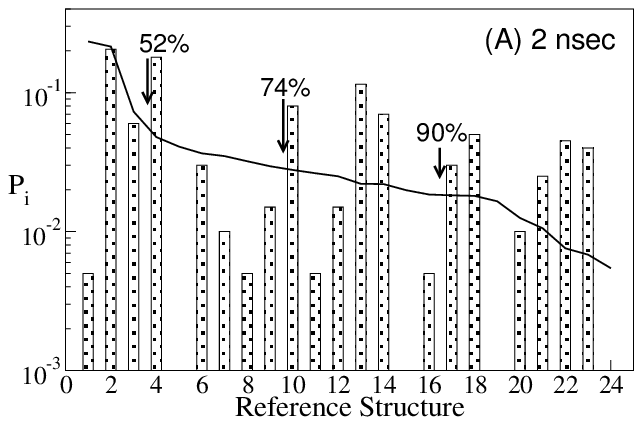,width = 0.50\columnwidth}
\epsfig{file=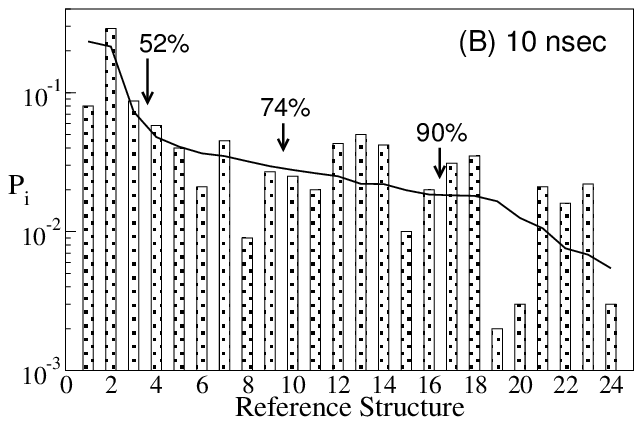,width = 0.50\columnwidth}
\epsfig{file=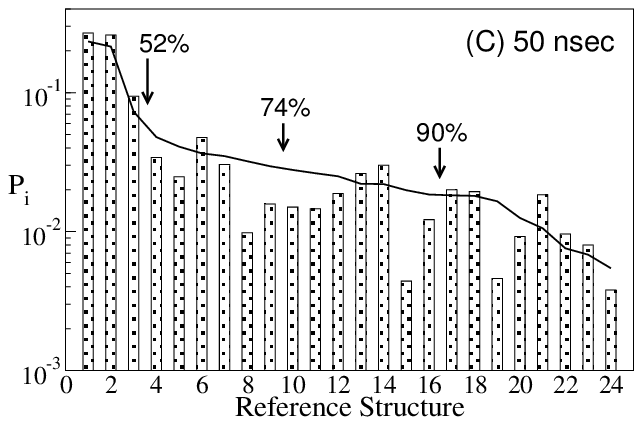,width = 0.50\columnwidth}
\epsfig{file=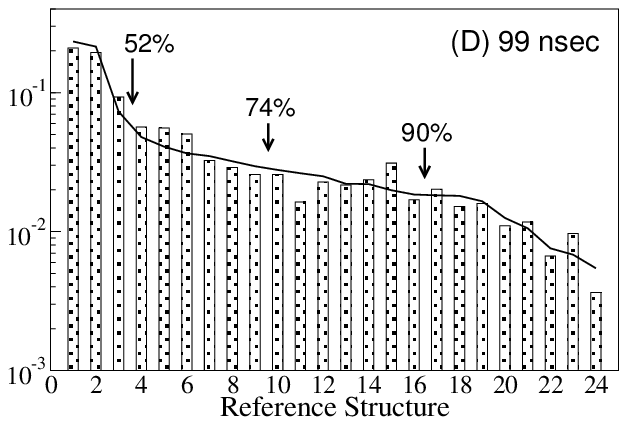,width = 0.50\columnwidth}
\caption{
}
\label{fig3}
\end{figure}

For met-enkephalin, we use our histogram approach to illustrate, in Fig.\ \ref{fig3}, 
the \emph{evolution} of convergence in two long ($99$ nsec) trajectories. The compound 
trajectory ($198$ nsec) is taken as a ``gold standard,'' from which reference structures 
are calculated using a cutoff $d_c=3.0$ \AA. We can then assess the convergence of portions of the
trajectory against this full ensemble (see Figs.\ \ref{fig3}A-D).  
%

From Fig.\ \ref{fig3}A, it is clear that after the first $2$ nsec, the 
simulation is far from converged. Many important substates have not 
yet been visited, and many of the bins are over or 
underpopulated by several $k_{B}T$. (On a semilog scale, a factor 
of $2$ in the population represents an error of $1/2$ $k_{B}T$.) 
After $50$ nsec (Fig.\ \ref{fig3}(C)), all clusters are populated, 
but many important substates have 
not converged to within $1/2$ $k_{B}T$ of the $198$ nsec values.

Fig.\ \ref{fig3} presents a picture of a very conformationally diverse 
peptide, especially given the large cutoff ($d_c=3.0$ \AA) used. 
The first $3$ ``substates'' contain only 52\% of the observed 
structures, while the first $9$ account for 74\%. Indeed, the 
(experimentally determined) starting structure is located in the 
second most populated bin. 

We also analyzed the entire set of NMR model structures. 
These were determined in the presence of bicelles, as it was hypothesized that interaction 
of the peptide with the cell membrane induces a shift in the conformational 
distribution\cite{metenk-bicelle-nmr}. We classified the entire set of $80$ NMR 
structures against our set of reference structures. The overwhelming majority of the 
NMR structures---$75$\%---were nearest to reference number $23$---the second-least populated 
bin in our simulation. The next largest group of NMR structures ($15$ of $80$) were nearest 
to reference number $2$, which held a comparable portion of the simulation trajectory. 
The remaining $5$ NMR structures were scattered among $4$ different 
bins. While not conclusive, the comparison between our simulation data and the NMR structures 
supports the hypothesis that binding to the membrane induces a shift in the distribution 
of met-enkephalin conformers, relative to the distribution observed in water. 
Such conformational diversity is not surprising for a 
peptide, which is known to be a promiscuous neurotransmitter 
by virtue of its flexibility\cite{metenk-book,metenk-nmr-flex,metenk-bicelle-nmr}. 
However, it will be interesting to revisit 
the issue in the study of a protein.
 
\subsubsection{Self-referential Convergence Assessment\label{convergencesec}}
We want to assess convergence \emph{without} the use of a ``gold-standard.'' Our 
previous analysis (Fig.\ \ref{fig3}) might be used to compare simulation 
protocols--ensembles from a new protocol may be compared to a ``gold-standard'' 
ensemble. (Here, the gold standard is the $198$ nsec compound trajectory.) 
However, it is not useful as a means of assessing the 
convergence of a single simulation. After all, given only a $4$ nsec trajectory, 
one would like an assessment without reference to ``the answer''. 

Fig.\ \ref{fig4} therefore demonstrates a purely self-referential scheme for "on the fly" analysis 
of a continuous trajectory. Fig.\ \ref{fig4}A compares, for example, the first $2$ nsec 
to the second $2$ nsec. The series of plots in Fig.\ \ref{fig4} shows that 
the populations of the clusters are still changing significantly, even 
between the first and second $50$ nsec. Presuming we had run only a single 
$100$ nsec simulation, we could make Fig.\ \ref{fig4}C, and describe the 
convergence by saying, ``at a resolution of $3.0$ \AA\  RMSD, considering bins 
containing $75\%$ of the structures, $6$ of $9$ bins have not yet converged 
to within $1/2$ $k_{B}T$.'' Note the contrast with Fig.\ \ref{fig2}B, 
where it appears convergence is reached after just $7$ nsec.    
This contrast is all the more striking considering that $d_{c}=3.0$ \AA\ 
is a rather conservative choice. At a higher resolution (smaller $d_{c}$) 
the observed convergence is worse.

To test whether our analysis is sensitive to the (random) selection 
of reference structures, Fig.\ \ref{fig4} shows two independent 
sets of reference structures. There is little difference in 
the results. Both classifications indicate that more than $50$ nsec 
are required for convergence when $d_{c}=3.0$ \AA.

The observed ensembles and corresponding convergence depend on 
both the metric used and the 
value of $d_{c}$. (This is of course true of any clustering 
algorithm.) It is therefore important to report this information 
along with any statements about the convergence of a particular 
simulation. Indeed, lowering the cutoff, and hence increasing the 
resolution of the classification, is bound to reduce the observed 
level of convergence. Instead of Fig.\ \ref{fig4}, in which each 
panel is a different length of the trajectory, we could 
have plotted the same trajectory length at different resolutions. 
At a high enough resolution, we will always 
find some substates which are under- or over-populated. 
In other words, since all trajectories are finite, a physically 
acceptable value of $d_{c}$ must be chosen.

While the choice of $d_{c}$ is somewhat \emph{ad hoc} in the present 
implementation, plots like those in Fig.\ \ref{fig4} still can 
provide valuable, quantitative information. For example, imagine that 
we wish to calculate the free energy difference between two 
experimentally known conformations, which differ 
by $3.0$ \AA\ RMSD. In this case, Fig.\ \ref{fig4} suggests that 
we cannot expect an accuracy better than $1/2$ $k_{B}T$. Perhaps 
more importantly, \emph{any} fixed choice of cutoff can be useful in comparing 
different simulation methods---even if the difficult question of 
absolute convergence is not addressed.      

%
%
\section{Discussion}
We have introduced a structure-based classification approach for the 
analysis of biomolecular simulation trajectories. 
The method provides a more rigorous evaluation of convergence 
than commonly used methods. Our approach is based on 
a simple intuitive picture---namely, a comparison of the relative populations 
of different conformational substates. The method is trivially applicable to 
simulations of proteins of any size. 

The results for met-enkephalin indicate that it takes quite some time 
($>50$ nsec) for the relative populations of the various substates to equilibrate, 
even with fairly promiscuous cutoff ($3.0$ \AA\ RMSD) which partitions the trajectory 
into relatively few bins. 
Because we can expect that many transitions 
into and out of each substate will be required to equilibrate their 
relative populations, a simple cluster counting approach (Fig.\ \ref{fig2}B) 
will present a deceptively optimistic picture of convergence.
In order to carefully assess convergence of a simulation, we must 
therefore compare the populations of the various substates from 
different fragments of the trajectory. A simple, fast way to carry out 
such a comparison is provided by the ensemble method 
described above. A higher level of rigor can be achieved by comparing  
multiple pairs of independent blocks of the trajectory.

It must be stressed that---though our method may provide an unambiguous 
\emph{negative} answer to the question, 
``is the simulation converged?''---it may only provide 
a \emph{provisionally positive} answer. A longer simulation may well reveal 
longer timescale phenomena, parts of structure space not yet visited. 

Our approach should be useful, in its present form, as a means to assess the 
\emph{relative} efficiencies of two simulation methods. (The cutoff $d_{c}$ can 
always be reduced enough to suggest  poorer convergence of at least one of 
the trajectories analyzed.) Many algorithms have recently 
generated broad interest by virtue of their potential to enhance 
the sampling of biomolecular conformation space. Some of these 
algorithms, notably the various parallel exchange 
simulations\cite{garcia-prl-big-repex-md}, 
invest considerable CPU time in pursuit of this goal. It is therefore 
important to ask whether these methods are in fact worth the extra 
expense, i.e.,``does running the algorithm in question increase 
the quantity: (observed conformational sampling)/(total CPU time)''? 

In 
particular, these parallel exchange algorithms should be compared 
to (i) single, parallelized trajectories, as are possible with NAMD\cite{namd}, for example, 
and (ii) multiple independent trajectories as suggested by 
Caves \emph{et.\ al}\cite{caves-multirun98}. 
The 
CPU time is easy enough to quantify, and we hope the present report will 
aid in evaluating the numerator.

In the future, we will study trajectories of larger proteins, in order 
to develop criteria for determining cutoffs in larger systems. On the one 
hand, the upper bound on RMSD distance between any pair of structures 
increases with the size of the protein. On the other hand, larger 
proteins may not be as structurally diverse as small, floppy 
peptides---at least on the timescale currently accessible to simulation. 
Work already underway on a G-protein coupled receptor should shed light 
on these issues\cite{alan-perscomm}. 
Furthermore, the approach should already be able to \emph{compare} 
different simulation methods in large systems. The systems which may be treated 
with our method are not limited to proteins, or even single chains. Indeed, the method is 
immediately applicable for analyzing simulations of polymers, nuclei acids, or 
macromolecular complexes. 

\clearpage

\begin{table}
\begin{center}
\begin{tabular}{|c|c|c|}
\hline $d_{c}$ in \AA\  & number of clusters & $\sigma$ \\ 
\hline $\mathbf{1.5}$ & $1860.0$ & $14.0$ \\ 
\hline $\mathbf{2.0}$ & $321.3$ & $6.7$\\
\hline $\mathbf{2.5}$ & $72.8$ & $3.8$\\ 
\hline $\mathbf{3.0}$ & $23.3$& $2.2$\\
\hline $\mathbf{3.5}$ & $10.3$ & $0.5$\\ \hline
\end{tabular}
\caption{ 
Average number of reference structures generated
for various cutoffs ($d_{c}$ in RMSD). Reported are the
average and standard deviation ($\sigma$) in the number of reference structures
for four independent clusterings of the plw-ab trajectory.
}
\label{cutoff}
\end{center}
\end{table}
\clearpage
\section*{Figure Legends}
\subsubsection*{Figure~\ref{fig2}.}
(A) RMSD from starting structure
for met-enkephalin trajectory plw-a.
(B) Number of populated
clusters vs. simulation time for the plw-a trajectory. Results are
shown for two independent clusterings. After
$7$ nsec, the simulation appears equilibrated. No more clusters appear in
the $198$ nsec plw-ab trajectory.
\subsubsection*{Figure~\ref{diffcutoffs}.}
Histograms for the plw-ab trajectory generated for
different values of $d_{c}$, indicated in the upper right corner
of each plot. $P_{i}$ is the normalized population of bin $i$, where
$i$ refers to the reference structure.
\subsubsection*{Figure~\ref{fig3}.}
\begin{figure}[bp]
\epsfig{file=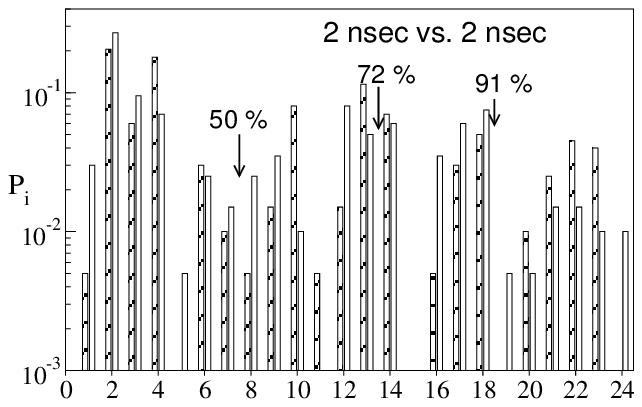,width = 0.50\columnwidth}
\epsfig{file=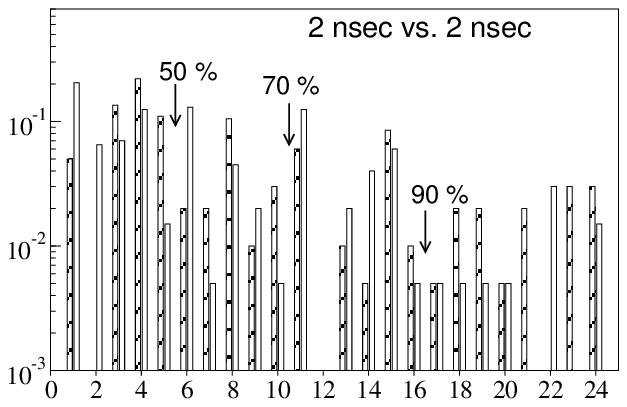,width = 0.50\columnwidth}
\epsfig{file=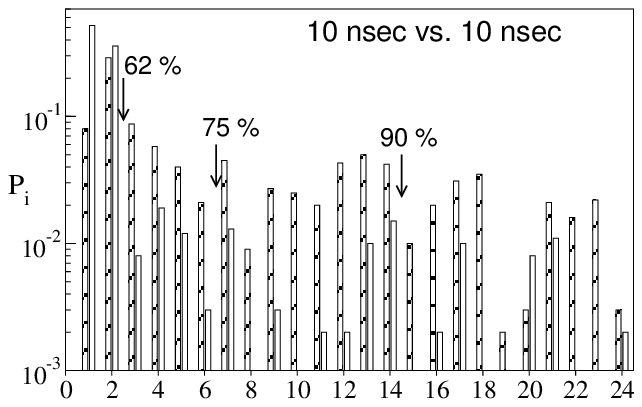,width = 0.50\columnwidth}
\epsfig{file=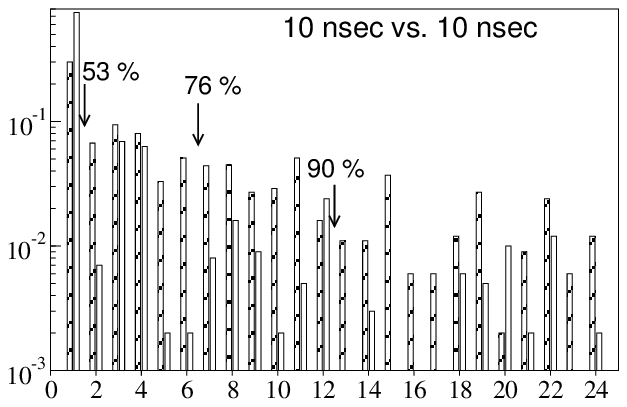,width = 0.50\columnwidth}
\epsfig{file=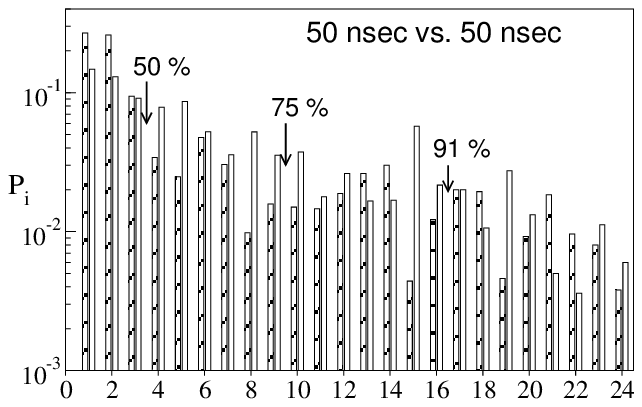,width = 0.50\columnwidth}
\epsfig{file=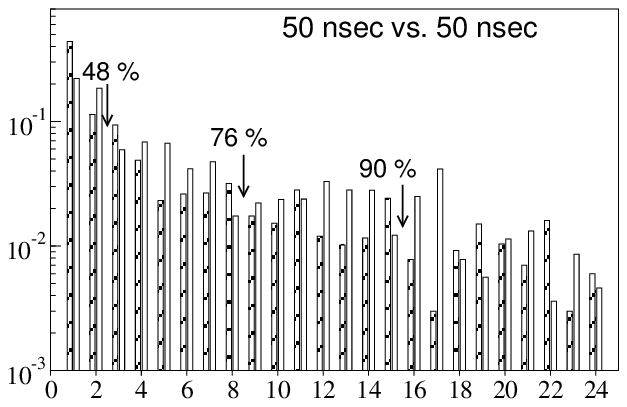,width = 0.50\columnwidth}
\epsfig{file=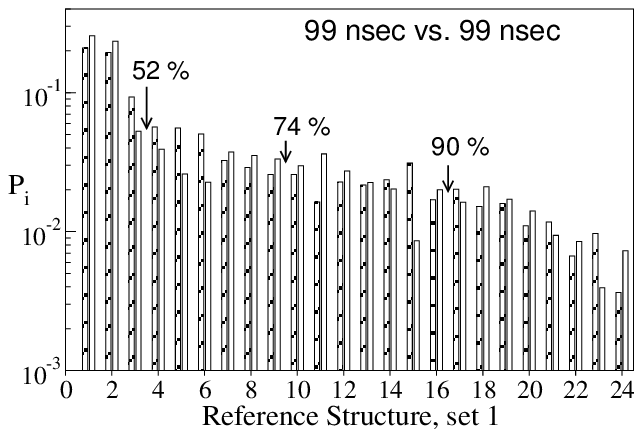,width = 0.50\columnwidth}
\epsfig{file=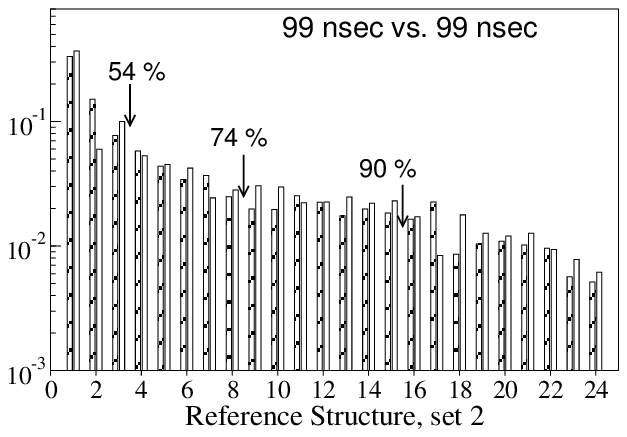,width = 0.50\columnwidth}
\caption{
}
\label{fig4}
\end{figure}

Ensembles for different fractions of trajectory plw-a
(bars), compared to the ensemble of the entire $198$ nsec compound trajectory
(solid line): $2$ nsec(A), $10$ nsec(B), $50$ nsec(C), $99$ nsec(D).
$d_{c}=3.0$ \AA\ RMSD. Note that $\text{ln}P_{i}$ is a
free-energy like quantity; hence on the semilog scale the difference
in populations may be read off in units of $k_{b}T$: a factor of $2$
on the y-axis corresponds to $0.5$ $k_{b}T$. The percentages indicate the
fraction of the $198$ nsec trajectory binned to that point.
\subsubsection*{Figure~\ref{fig4}.}
Self-consistent convergence of different trajectory lengths for two
independent classifications (``set $1$''and ``set $2$'')of the plw-ab
trajectory at $d_{c}=3.0$ \AA. Each
plot compares the first half (diagonal fill) to the second half
(gray shading) of the trajectory for total trajectory lengths of
(A), $4$ nsec; (B), $20$ nsec; (C), $100$ nsec; (D), $198$ nsec.
Percentages indicate the portion of the total trajectory
binned to that point.
\clearpage
\clearpage

\clearpage

\clearpage

\end{document}